\documentclass[conference]{IEEEtran}
\IEEEoverridecommandlockouts
\usepackage{cite}
\usepackage{amsmath,amssymb,amsfonts, amsthm}
\usepackage{xspace}
\usepackage[]{hyperref}
\usepackage{colortbl}
\usepackage{algorithm}
\usepackage{algpseudocode}
\usepackage{tikz}
\usetikzlibrary{positioning}
\usetikzlibrary{shapes.geometric}
\usepackage{listings}
\usepackage{xcolor}
\usepackage{minted}
\usepackage{tabularx}
\usepackage[most]{tcolorbox}
\usepackage{multirow}
\usepackage{subfig}
\usepackage{soul}
\usepackage{enumitem}

\newcommand{\tool}{\textsc{Hermes}\xspace}
\floatname{algorithm}{Procedure}
\algnewcommand{\MyComment}[1]{{\fontsize{8pt}{8pt}\selectfont\Comment{\textcolor{darkgray!80}{#1}}}}

\definecolor{codegreen}{rgb}{0,0.6,0}
\definecolor{codegray}{rgb}{0.5,0.5,0.5}
\definecolor{codepurple}{rgb}{0.58,0,0.82}
\definecolor{backcolour}{rgb}{0.95,0.95,0.92}

\lstdefinestyle{mystyle}{
  backgroundcolor=\color{backcolour}, 
  commentstyle=\color{codegreen},
  keywordstyle=\color{blue},
  numberstyle=\tiny\color{black},
  stringstyle=\color{codepurple},
  basicstyle=\ttfamily\footnotesize,
  breakatwhitespace=false,         
  breaklines=true,                 
  captionpos=b,                    
  keepspaces=true,                 
  numbers=left,                    
  numbersep=5pt,                  
  showspaces=false,                
  showstringspaces=false,
  showtabs=false,                  
  tabsize=2
}

\newif\ifshowcomments
\showcommentstrue

\ifshowcomments

\def\BibTeX{{\rm B\kern-.05em{\sc i\kern-.025em b}\kern-.08em
    T\kern-.1667em\lower.7ex\hbox{E}\kern-.125emX}}

\begin{document}

\title{An Empirical Study of Cross-Language Interoperability in Replicated Data Systems}

\author{
\IEEEauthorblockN{Provakar Mondal and Eli Tilevich}
\IEEEauthorblockA{\textit{Software Innovations Lab, Virginia Tech, USA}\\
\{provakar, tilevich\}@cs.vt.edu}\\}

\maketitle
\thispagestyle{plain}
\pagestyle{plain}

\begin{abstract}
BACKGROUND: 
Modern distributed systems replicate data across multiple execution sites. Business requirements and resource constraints often necessitate mixing different languages across replica sites. To facilitate the management of replicated data, modern software engineering practices integrate special-purpose replicated data libraries (RDLs) that provide read-write access to the data and ensure its synchronization. Irrespective of the implementation languages, an RDL typically uses a single language or offers bindings to a designated one. Hence, integrating existing RDLs in multilingual environments requires special-purpose code, whose software quality and performance characteristics are poorly understood. 

AIMS: We aim to bridge this knowledge gap to understand the software quality and performance characteristics of RDL integration in multilingual environments. 

METHOD: We conduct an empirical study of two key strategies for integrating RDLs in the context of multilingual replicated data systems: foreign-function interface (FFI) and a common data format (CDF); we measure and compare their respective software metrics and performance to understand their suitability for the task at hand.

RESULTS: Our results reveal that adopting CDF for cross-language interaction offers software quality, latency, memory consumption, and throughput advantages. We further validate our findings by (1) creating a CDF-based RDL for mixing compiled, interpreted, and managed languages; and (2) enhancing our RDL with plug-in extensibility that enables adding functionality in a single language while maintaining integration within a multilingual environment. 

CONCLUSIONS: With modern distributed systems utilizing multiple languages, our findings provide novel insights for designing RDLs in multilingual replicated data systems.
\end{abstract}


\section{Introduction} \label{sec:intro}
In distributed systems, data is often replicated across multiple execution sites, known as \emph{replicas}, to lower access latency, increase data availability, and prevent a single point of failure~\cite{eischer2020resilient}. As replicas modify their local state, the updates must be synchronized across replicas to keep them consistent, accomplished via a consistency protocol. Modern distributed systems commonly deploy replicated data libraries (RDLs) to manage all interactions with and synchronization of data. 

RDL provides API methods for accessing and updating replicated data, synchronizing the updates behind the scenes~\cite{burckhardt2014replicated}. A local operation or a remotely initiated runtime synchronization request can update a replica's state. RDL's programming interface hides away the complexity of its runtime while ensuring consistency of the replicated state~\cite{jagadeesan2018eventual}. RDL examples include Conflict-free Replicated Data Types (CRDTs)~\cite{Shapiro2011CRDTs}, Explicitly Consistent Replicated Objects (ECROs)~\cite{de2021ecros}, and Mergeable Replicated Data Types (MRDTs)~\cite{kaki2019mergeable}.

The need to replicate data in modern distributed applications has led to the creation of various RDLs, most of which are implemented in a single programming language~\cite{de2020cscript}. If written in the same language, application code and an RDL can interact straightforwardly. When the business needs demand that application code at different replicas be implemented in dissimilar languages, developers can follow two strategies for integrating an RDL~\cite{mondal2025understanding, bisiani1987architectural, li2024multilingual}. \textbf{Strategy I}: the application code interfaces with a monolingual RDL via the foreign function interfaces (FFIs). \textbf{Strategy II}: the RDL supports multiple languages; so the application code interacts with the library natively, while the replicas synchronize their state by exchanging messages in a common data format (CDF). Although some RDLs provide bindings for cross-language interoperability~\cite{crdt_languages} for specific language pairs (e.g., Rust$\leftrightarrow$JavaScript~\cite{loro2024crdt}), one can classify them as special-purpose FFIs. 
 
For the challenging task of mixing languages within replicated data applications, it remains unclear which of the two strategies mentioned above should be used, as their software quality and performance characteristics remain unexplored. Prior research has pointed out how FFI can impose the complexity of additional dependent libraries and convoluted build procedures~\cite{turcotte2019reasoning}. However, FFI usage has not been studied in the context of replicated data systems, nor has it been compared with alternatives in this domain. 

To address this knowledge gap, this paper empirically studies Strategies I and II, assessing how their software quality and performance characteristics influence their suitability for overcoming language barriers in this domain. Our results demonstrate that Strategy II offers software quality and performance advantages. Hence, we further validate these findings by creating our own multilingual RDL, implemented as \tool\footnote{Our system's title is inspired by the Greek mythological god Hermes, with the ability to speak different languages, as he is the god of interpreters, translators, and communication~\cite{hermes}.}. \tool facilitates cross-language coordination of replicas across compiled, interpreted, and managed languages. With \tool, developers can choose the most appropriate languages for distributed replicas, as guided by the business logic requirements.

Due to evolving business requirements, RDLs may need to offer additional functionalities, such as persistence and fault tolerance~\cite{saquib2022log, kleppmann2022making}. These distributed features often require integration with library code written in various programming languages. Implementing each additional RDL feature separately in all supported languages would be inefficient. To alleviate this complication, \tool provides plug-in support for systematic extension with extra features. Developers can implement a \tool plug-in in a single language, and the resulting feature will be seamlessly integrated with replicas across different languages. To that end, \tool takes declarative metadata as input and generates the required cross-language coordination functionality. In this way, \tool bridges language barriers by providing core RDL features in multilingual environments and offering a structured approach for extending libraries with additional functionalities.

This paper's contributions are: 
\begin{enumerate}
    \item An empirical study of two key strategies for RDL integration in multilingual replicated data systems.
    \item An RDL design based on a common data format (CDF) for cross-language interaction, with plug-in extensibility for additional features.
    \item A reference implementation---\tool, featuring replicated Counter, Set, and Map data types in compiled (Go), interpreted (JavaScript), and managed (Java) languages, as well as a validation of its plug-in extensibility that highlights the ability to integrate complex features.
\end{enumerate}

The rest of this paper is structured as follows: Section \ref{sec:back-motivation} describes the background and motivates the empirical study; Section \ref{sec:empirical-study} presents our empirical study; Section \ref{sec:system} details \tool's system design and workflow; Section~\ref{sec:implementation} describes \tool's implementation including its plug-in extensibility; Section \ref{sec:related} presents the related state of the art; and Section \ref{sec:conclusion} concludes the paper with future work plans.


\section{Background \& Motivation} \label{sec:back-motivation}
This section presents the technical background, including general concepts and specific technologies. Then we present a software development scenario that motivates this research.

\subsection{Background}
\subsubsection{Programming Frameworks for Multilingual Distributed Systems}
Numerous approaches have been introduced to support the development of multilingual distributed systems. In Common Object Request Broker Architecture (CORBA), the Interface Definition Language (IDL) is used to define interfaces implemented in multiple languages~\cite{vinoski1997corba}. CORBA's inter-language Remote Procedure Call (RPC) marshals arguments/return types, and maps errors reporting to language-specific constructs. Java-based MapReduce provides language bindings to Python and C++~\cite{dean2008mapreduce}. MapReduce open-source implementation, Hadoop, supports additional languages by providing language-specific marshaling and input/output~\cite{borthakur2007hadoop}. Spark extends multilingual support to Java, Scala, Python, and R by relying on JVM services~\cite{zaharia2010spark}. In the RESTful architecture, heterogeneous peers communicate via a limited number of verbs, such as HTTP GET, PUT, POST, etc~\cite{richardson2008restful}. 


\subsubsection{Replicated Data Libraries (RDLs)}
Having briefly introduced RDLs earlier, we next describe their basic characteristics. A typical RDL consists of three major components: (1) \texttt{state}, (2) \texttt{interface}, and (3) \texttt{update propagation}~\cite{mao2022reversible}. The \texttt{state}, the internal data structure, represents the data type; RDL data types range from basic data structures, including counter, set, and map to more complex ones, such as a JSON document~\cite{kleppmann2017conflict}. The \texttt{interface} provides an API, a set of operations for clients to access and update the RDL state. The \texttt{update propagation} refers to the procedure that propagates local updates to the remaining replicas ensuring their convergence. 




\subsubsection{Plug-in Extensibility}
\begin{figure}[hbt!]
    \centering
    \begin{tikzpicture}
        \node [draw, rectangle, minimum width=2.4cm, minimum height=0.6cm, inner sep=0] at (0,0) (PL) {\footnotesize{Plug-in Loader}};
        \node [draw, rectangle, fill=gray!10, minimum width=2.4cm, minimum height=0.2cm, inner sep=0, below = 0pt of PL] (PLB) {};
        \node [draw, rectangle, minimum width=2.4cm, minimum height=1cm, inner sep=0, below = 0pt of PLB] (PLF) {\scriptsize{+getPlugIns():PlugIn[]}};

        \node [draw, rectangle, minimum width=2.2cm, minimum height=0.6cm, inner sep=0] at (3,0) (PI) {\footnotesize{Plug-in Interface}};
        \node [draw, rectangle, fill=gray!10, minimum width=2.2cm, minimum height=0.2cm, inner sep=0, below = 0pt of PI] (PIB) {};
        \node [draw, rectangle, minimum width=2.2cm, minimum height=1cm, inner sep=0, below = 0pt of PIB] (PIF) {\scriptsize{\begin{tabular}{@{}l@{}}
                        +func\textsubscript{1}: int \\
                        +func\textsubscript{2}: string \\
                        ...\\
                        \end{tabular}}};

        \node [draw, rectangle, minimum width=2.2cm, minimum height=0.6cm, inner sep=0] at (6,0) (PC) {\footnotesize{Concrete Plug-in}};
        \node [draw, rectangle, fill=gray!10, minimum width=2.2cm, minimum height=0.2cm, inner sep=0, below = 0pt of PC] (PCB) {};
        \node [draw, rectangle, minimum width=2.2cm, minimum height=1cm, inner sep=0, below = 0pt of PCB] (PCF) {\scriptsize{\begin{tabular}{@{}l@{}}
                        +func\textsubscript{1}: int \\
                        +func\textsubscript{2}: string \\
                        ...\\
                        \end{tabular}}};

        \draw[densely dashed, -, line width=1pt] (PCB.west) -- (PIB.east);

        \draw[->, line width=1pt] (PLB.east) -- (PIB.west);
    \end{tikzpicture}
    \caption{A Typical Plug-in Structure}
    \vspace{-0.4cm}
    \label{fig:plug-in-pattern}
\end{figure}
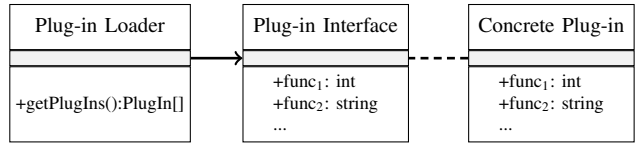

Sometimes, developers need to add new features to an application without modifying its core functionality. A common approach to accomplish this objective is the plug-in design pattern~\cite{birsan2005plug}. The known advantages of this pattern include extensibility, flexibility, and isolation of core application features from custom logic. A high-level structure of the plug-in design pattern appears in Figure~\ref{fig:plug-in-pattern}. It depicts this pattern's three key components: (1) Plug-in Loader, (2) Plug-in Interface, and (3) Concrete Plug-in~\cite{mayer2003lightweight}. Plug-in Loader is responsible for locating a plug-in interface and ensuring that the corresponding plug-in modules are loaded and ready for invocation; the diagram depicts these functionalities via the \lstinline{getPlugIns()} function. Plug-in Interface exposes the method interface to invoke the functions provided by a plug-in. Concrete Plug-in represents the actual plug-in module provided as an external feature to be integrated with the main application. Plug-ins have been widely applied as a systematic mechanism for enhancing core functionalities with extra features in domains that include web services, content management systems, and integrated development environments~\cite{devleader2023plugin}. 

\subsection{Motivation}
Next, we present a software development scenario that motivates the need to mix languages in a replicated data system. Using this example, we identify the need for a deeper understanding of strategies for RDL integration.

\subsubsection*{Motivating Example} \label{sub:motivating-example}
\begin{figure}[hbt!]
    \centering
    \includegraphics[scale=0.7]{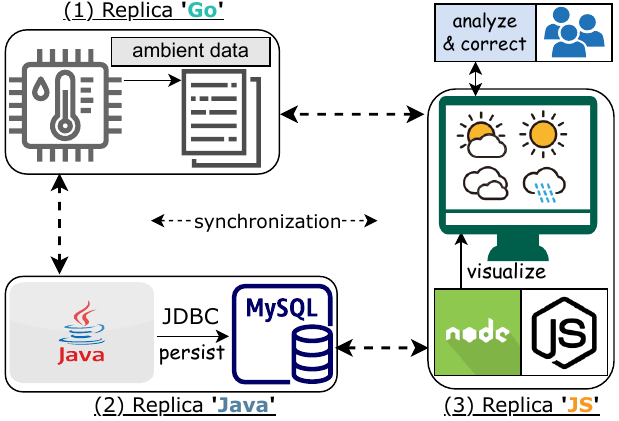}
    \caption{Ambient Data: Collect, Persist, and Visualize}
    \label{fig:example}
    \centering
\end{figure}

Figure \ref{fig:example} depicts a distributed system that manages ambient data, such as air temperature and humidity. The system comprises three replica nodes: (1) a collection node, hosting an IoT device that collects the data; (2) a database node, hosting a database engine; and (3) a monitoring node, that graphically depicts the collected data, so human analysts can examine and correct it if necessary. Each replica is implemented in a language best suited for its functionality: (1) the IoT device runs Go to capture sensor data resource-efficiently (\texttt{Replica Go}); (2) the database engine uses a popular JDBC API~\cite{fisher2003jdbc} for easy database interfacing (\texttt{Replica Java}); and (3) the monitoring functionality uses the powerful Node.js graphical frameworks (\texttt{Replica JS}). Each programming language and its standard library are known for particular suitability for different tasks, one of the reasons for the very co-existence of numerous programming languages~\cite{stefik2014programming}. The data, replicated across these three replicas can be expressed as an RDL. As ambient data is collected, the updates propagate to the persistence (\texttt{Replica Java}) and monitoring (\texttt{Replica JS}) replicas; if an analyst corrects the data, the updates also synchronize with the remaining replicas. 

Despite the wealth of existing RDLs, none can be used \emph{as is} in this example, with three replicas each using a different language. A monolingual RDL would be inapplicable; an RDL with a binding to a designated language would be inapplicable either unless the example uses that particular language as well. Recall that in Section \ref{sec:intro}, we discuss two strategies for structuring cross-language interoperability for multilingual replicas. However, which of these strategies should be preferred remains unknown without solid evidence about their respective software quality and performance characteristics. For example, would employing FFI for applications to interoperate with RDL substantially complicate the resulting software architecture? Would the resource constraints of the \texttt{Replica Go} preclude meeting the memory consumption requirements of the FFI libraries? 

This motivating example informs the empirical study of the two aforementioned integration strategies we have conducted. Even though we could have evaluated various aspects of these strategies, we focus on the facets of their software structure and execution that can affect the application scenarios as exemplified in this motivating example. Specifically, this example mixes compiled, interpreted, and managed languages to leverage their respective strength in order to accomplish particular requirements. Despite its specific application use case, this example is representative of numerous systems that employ replicated data to fulfill their business requirements. 

\section{Empirical Study} \label{sec:empirical-study}
The following research questions drive our empirical study:
\begin{enumerate}
    \item \textbf{RQ1: Software Quality:} Which of the integration strategies possesses superior software quality, as evidenced by measures of programming burden, Cyclomatic complexity, and Halstead effort?
    \item \textbf{RQ2: Performance Characteristics:} Which of the strategies offers superior performance in terms of latency, memory consumption, and throughput?
\end{enumerate}

We start by describing the two reviewed strategies for cross-language interoperability. Then we present the results of our software quality and performance experiments, which answer the research questions above. Finally, we discuss those findings that we chose to further validate by implementing our CDF-based RDL, \tool.

\subsection{Reviewed Strategies for Cross-Language Interoperability}
\begin{figure}[hbt!]
    \centering
    \begin{tikzpicture}
        \node [draw, rectangle, minimum width=2.8cm, minimum height=2.1cm, inner sep=0, 
                rounded corners, densely dashed, line width=0.8pt] at (0,0) (R1) {};
        \node at ([yshift=-0.2cm]R1.north) {\footnotesize{Replica R\textsubscript{1}}};
        \node [draw, rectangle, minimum width=2.4cm, minimum height=0.5cm, inner sep=0] at ([yshift=-0.7cm]R1.north) (App1) {\footnotesize{App-Language: L\textsubscript{1}}};
        \node [draw, rectangle, minimum width=2.4cm, minimum height=0.5cm, inner sep=0] at ([yshift=0.4cm]R1.south) (CRDT1) {\footnotesize{RDL-Language: L\textsubscript{x}}};
        \draw[<->, thick] ([xshift=-8pt]App1.south) -- ([xshift=-8pt]CRDT1.north) node[midway, right] {\footnotesize{FFI \textcolor{violet}{(L\textsubscript{1}$\leftrightarrow$L\textsubscript{x})}}};

        \node [draw, rectangle, minimum width=2.8cm, minimum height=2.1cm, inner sep=0, 
                rounded corners, densely dashed, line width=0.8pt] at (5.2,0) (R2) {};
        \node at ([yshift=-0.2cm]R2.north) {\footnotesize{Replica R\textsubscript{2}}};
        \node [draw, rectangle, minimum width=2.4cm, minimum height=0.5cm, inner sep=0] at ([yshift=-0.7cm]R2.north) (App2) {\footnotesize{App-Language: L\textsubscript{2}}};
        \node [draw, rectangle, minimum width=2.4cm, minimum height=0.5cm, inner sep=0] at ([yshift=0.4cm]R2.south) (CRDT2) {\footnotesize{RDL-Language: L\textsubscript{x}}};
        \draw[<->, thick] ([xshift=-8pt]App2.south) -- ([xshift=-8pt]CRDT2.north) node[midway, right] {\footnotesize{FFI \textcolor{magenta}{(L\textsubscript{2}$\leftrightarrow$L\textsubscript{x})}}};

        \draw[<->] (CRDT1.east) -- (CRDT2.west) node[midway, above] {\footnotesize{synchronization}};;
        
        \node[align=center] at (3, -1.5) {\footnotesize{Strategy 1: FFI for Cross-Language Interoperability}}; 
        
        \draw[solid] (-1.6, -2) -- (6.8, -2);
        
        \node [draw, rectangle, minimum width=3.2cm, minimum height=2.1cm, inner sep=0, 
                rounded corners, densely dashed, line width=0.8pt] at (0,-3.5) (R1N) {};
        \node at ([yshift=-0.2cm]R1N.north) {\footnotesize{Replica R\textsubscript{1}}};
        \node [draw, rectangle, minimum width=2.4cm, minimum height=0.5cm, inner sep=0] at ([yshift=-0.7cm]R1N.north) (App1N) {\footnotesize{App-Language: L\textsubscript{1}}};
        \node [draw, rectangle, minimum width=2.4cm, minimum height=0.5cm, inner sep=0] at ([yshift=0.4cm, xshift=-9pt]R1N.south) (CRDT1N) {\footnotesize{RDL-Language: L\textsubscript{1}}};
        \node [draw, diamond, minimum height=0.3cm, minimum width=0.6cm, inner sep=0] at ([xshift=1.5cm]CRDT1N.center) (CDF1) {\scriptsize{CDF}};
        \draw[<->, thick, color=teal] ([xshift=-9pt]App1N.south) -- (CRDT1N.north) node[midway, right] {\footnotesize{Native}};

        \node [draw, rectangle, minimum width=3.2cm, minimum height=2.1cm, inner sep=0, 
                rounded corners, densely dashed, line width=0.8pt] at (5.2,-3.5) (R2N) {};
        \node at ([yshift=-0.2cm]R2N.north) {\footnotesize{Replica R\textsubscript{2}}};
        \node [draw, rectangle, minimum width=2.4cm, minimum height=0.5cm, inner sep=0] at ([yshift=-0.7cm]R2N.north) (App2N) {\footnotesize{App-Language: L\textsubscript{2}}};
        \node [draw, rectangle, minimum width=2.4cm, minimum height=0.5cm, inner sep=0] at ([yshift=0.4cm, xshift=9pt]R2N.south) (CRDT2N) {\footnotesize{RDL-Language: L\textsubscript{2}}};
        \node [draw, diamond, minimum height=0.3cm, minimum width=0.6cm, inner sep=0] at ([xshift=-1.5cm]CRDT2N.center) (CDF2) {\scriptsize{CDF}};
        \draw[<->, thick, color=teal] (App2N.south) -- ([xshift=-9pt]CRDT2N.north) node[midway, right] {\footnotesize{Native}};

        \draw[<->] (CDF1.east) -- (CDF2.west) node[midway, above] {\footnotesize{synchronization}};;
        
        \node[align=center] at (3, -5) {\footnotesize{Strategy 2: CDF for Cross-Language Interoperability}}; 
    \end{tikzpicture}
    \caption{Cross-Language Integration Strategies}
    \vspace{-0.2cm}
    \label{fig:multilingual_archs}
\end{figure}

Figure~\ref{fig:multilingual_archs} depicts the two strategies for integrating RDLs in multilingual replicated data systems. In both strategies, we assume that the system comprises two replicas: R\textsubscript{1} and R\textsubscript{2}. The application logic in R\textsubscript{1} is written in language L\textsubscript{1}, while in R\textsubscript{2} it is written in  L\textsubscript{2}. In Strategy I at the top, replicated data management is provided by a monolingual RDL, written in language L\textsubscript{x}. In this strategy, the application interacts with the library via FFI, whose type is determined by the specific pair of interacting languages. In this case, two different FFIs are required: L\textsubscript{1}$\leftrightarrow$L\textsubscript{x} for replica R\textsubscript{1}, and L\textsubscript{2}$\leftrightarrow$L\textsubscript{x} for replica R\textsubscript{2}. Please note that the direction of language interaction can further differentiate the type of FFI libraries required~\cite{hu2023cross}. In other words, FFI for L\textsubscript{1}$\rightarrow$L\textsubscript{x} may not be applicable for L\textsubscript{x}$\rightarrow$L\textsubscript{1}. 

In Strategy II at the bottom, the RDL is multilingual, meaning that for replica R\textsubscript{1}, the library code is written in L\textsubscript{1}, while in L\textsubscript{2} for replica R\textsubscript{2}. That is, the language of the RDL is the same as that of the replica's application logic. Hence, the application code interacts natively with the library through regular local function calls. However, when exchanging messages across replicas, the message format needs to be mapped to a common data format, marked as CDF in the figure. 

Modern systems widely utilize both strategies~\cite{uber_marketplace_ffi, salesforce_coding}. However, to the best of our knowledge, these strategies have never been compared side-by-side to understand their tradeoffs, a shortcoming this work aims to address.

\subsection{Results} \label{subsec:evaluation}
We first describe our experimental setup and evaluation subjects, and conclude with our results.

\begin{table*}[hbt!]
    \centering
    \caption{Software Quality Metrics: Lower Scores Indicate Higher Quality}
    \resizebox{\textwidth}{!}{
    \begin{tabular}{|l|c|c|c||c|c|c||c|c|c|}
        \hline
        \multirow{2}{*}{Libraries} & \multicolumn{3}{c||}{Extra ULOC} & \multicolumn{3}{c||}{Cyclomatic Complexity} & \multicolumn{3}{c|}{Halstead Effort}\\ \cline{2-10} 
        & Go & JavaScript & Java & Go & JavaScript & Java & Go & JavaScript & Java \\ \hline \hline
        Strategy I (Go-CRDT) & None & 358 & 406 & 28 & 46 & 102 & 5784.9 & 18915.75 & 45281.67 \\ \hline
        Strategy I (Legion) & 214 & None & 345 & 44 & 33 & 51 & 71890.7 & 9450.2 & 93057.5 \\ \hline
        Strategy I (T-CRDT) & 231 & 165 & None & 64 & 59 & 42 & 69758.01 & 80271.35 & 7479.97 \\ \hline \hline
        Strategy II (CDF-RDL) & \textbf{None} & \textbf{None} & \textbf{None} & \textbf{28} & \textbf{33} & \textbf{42} & \textbf{5784.9} & \textbf{9450.2} & \textbf{7479.97} \\ \hline
    \end{tabular}}
    \label{table:se-quality}
\end{table*}

\subsubsection*{Experimental Setup}
For our experiments, we implemented the system used as our motivating example described in Section~\ref{sub:motivating-example}: a system that collects, persists, and visualizes ambient weather data, replicated across three distributed sites. We deployed our implementation on a three-replica cluster with the following configurations. \texttt{Replica Go} is hosted on a 32-bit Raspbian 9 Raspberry Pi 3 with a DHT11 temperature-humidity sensor that collects ambient data from the surrounding environment. On this replica, we used cgroups~\cite{cgroups}, a Linux kernel feature that limits the available CPU processing power, thus imitating a genuine resource-constrained environment. \texttt{Replica Java} is hosted on a 64-bit Ubuntu 20.04 desktop with the application that persists the freshest data in MySQL via JDBC. \texttt{Replica JS} is hosted on a 64-bit Ubuntu 20.04 laptop running Node.js to visualize the data. This replica provides facilities for users to view, analyze, and correct the ambient data. This example uses three replicated data types: Counter to track the total collected data points; Set to maintain the uniqueness of the data points in the system; and Map to store key-value pairs representing the mapping from the collection timestamp to the collected data. 

\subsubsection*{Evaluation Subjects}
To evaluate Strategy I, we experimented with three open-source third-party RDLs. Each library is monolingual, written in one of the languages of the motivating example. Go-CRDT~\cite{go_crdt} is Go's built-in RDL package, available from version 1.19. Legion~\cite{legion} provides a JavaScript implementation of various replicated data types and a WebRTC-based networking layer. T-CRDT~\cite{t-crdt} is a Java RDL that allows programmers to either choose one of the predefined concurrency-conflict policies or specify their own. From each RDL, we used Counter, Set, and Map. For these monolingual libraries to work with our example application, we provided FFIs to each library's API. For example, Go-CRDT works directly with \texttt{Replica Go}; however, to use it with \texttt{Replica Java} and \texttt{Replica JS}, we had to route the library interactions via Go$\leftrightarrow$Java and Go$\leftrightarrow$JavaScript FFIs, respectively. Similarly, for Legion, we routed the interactions via JavaScript$\leftrightarrow$Go and JavaScript$\leftrightarrow$Java FFIs. Finally, for T-CRDT, we routed the library interactions via Java$\leftrightarrow$Go and Java$\leftrightarrow$JavaScript FFIs to make these RDLs work with our application. Hence, although these RDLs are monolingual, our multilingual replicas access the required RDL API via FFI.

To evaluate Strategy II, we implemented a multilingual RDL in Go, JavaScript, and Java, enabling communication between these languages through a common data format (CDF). For our CDF, we chose Google's Protocol Buffer (protobuf)~\cite{protobuf}, one of the most recent implementations of this concept supported by a major corporation. In particular, protobuf provides language- and platform-neutrality with its extensible data serialization~\cite{hatledal2019language}. Further, protobuf's compact binary message format reduces the amount of transmitted data, thus potentially improving network latency and throughput, especially in high-performance or resource-constrained environments~\cite{schwitzer2011using}. 


\subsection*{RQ1: Software Quality}
\begin{tcolorbox}[colback=gray!10!white, colframe=black, arc=1.5mm, boxrule=0.8pt]
\textbf{Summary:} Across all measurements, Strategy II is superior to Strategy I, based on the programming burden of integrating an RDL, Cyclomatic complexity, and Halstead effort.
\end{tcolorbox}

To assess the additional programming burden for the application logic to interface with the RDLs, we measured the uncommented lines of code (ULOC) required for it. ULOC has been used as an estimate of the programming effort~\cite{morozoff2009using}. The more code the application programmer has to write to integrate an RDL with application code, the higher the resulting programming burden is. To assess how complex the integration code is and how difficult it is to understand and maintain, we measured the Cyclomatic complexity and Halstead effort metrics. These metrics are among the most commonly used in assessing software quality and the maintenance effort required. In particular, Cyclomatic complexity measures the number of linearly independent paths that go across the source code~\cite{gill1991cyclomatic}. By counting the decision points (i.e., conditionals, loops, etc.), this metric embodies the source code's complexity. The more paths there are, the higher the resulting Cyclomatic complexity is. Halstead effort, a part of the Halstead metrics, derives how complex the source code is from its number of operators and operands~\cite{hariprasad2017software}. This metric assumes that the presence of operators and operands affects the cognitive burden of comprehending the source code. For both Cyclomatic complexity and Halstead effort, lower values indicate higher software quality~\cite{alfadel2017evaluation}.

All three software metrics appear in Table \ref{table:se-quality}. Our measurements focus on the integration functionality: the application logic invoking RDL API. ULOC measures the amount of additional code application programmers have to write, while Cyclomatic complexity and Halstead effort measure the quality of all the code of that functionality. Because in Strategy II, each supported language invokes the RDL's API methods natively, integration functionality requires no additional code. Furthermore, Strategy II exhibits lower Cyclomatic complexity and Halstead effort values, indicating that the integration functionality is simpler, making it easier to understand and maintain than the FFI-based alternatives.

For each language, Strategy II follows the same implementation as that of the corresponding native RDL. Hence, the RDL API functions are accessed in identical ways. Consequently, the Cyclomatic complexity and Halstead Effort measurements in Strategy II are the same as in Strategy I's corresponding native language RDL. For example, for Strategy I (T-CRDT in Java), these measurements for Java are 42 and 7479.97, respectively. As the Java implementation of Strategy II follows the same logic as Strategy I's T-CRDT, these values for Java remain the same.

\subsection*{RQ2: Performance Characteristics}
\begin{tcolorbox}[colback=gray!10!white, colframe=black, arc=1.5mm, boxrule=0.8pt]
\textbf{Summary:} Across all experiments, Strategy II exhibits shorter latency, smaller memory consumption, and higher throughput than Strategy I.
\end{tcolorbox}

\begin{figure*}[hbt!]
    \centering
    \subfloat[Counter\label{subfig:counter_latency}]
    {\includegraphics[width=.33\linewidth]{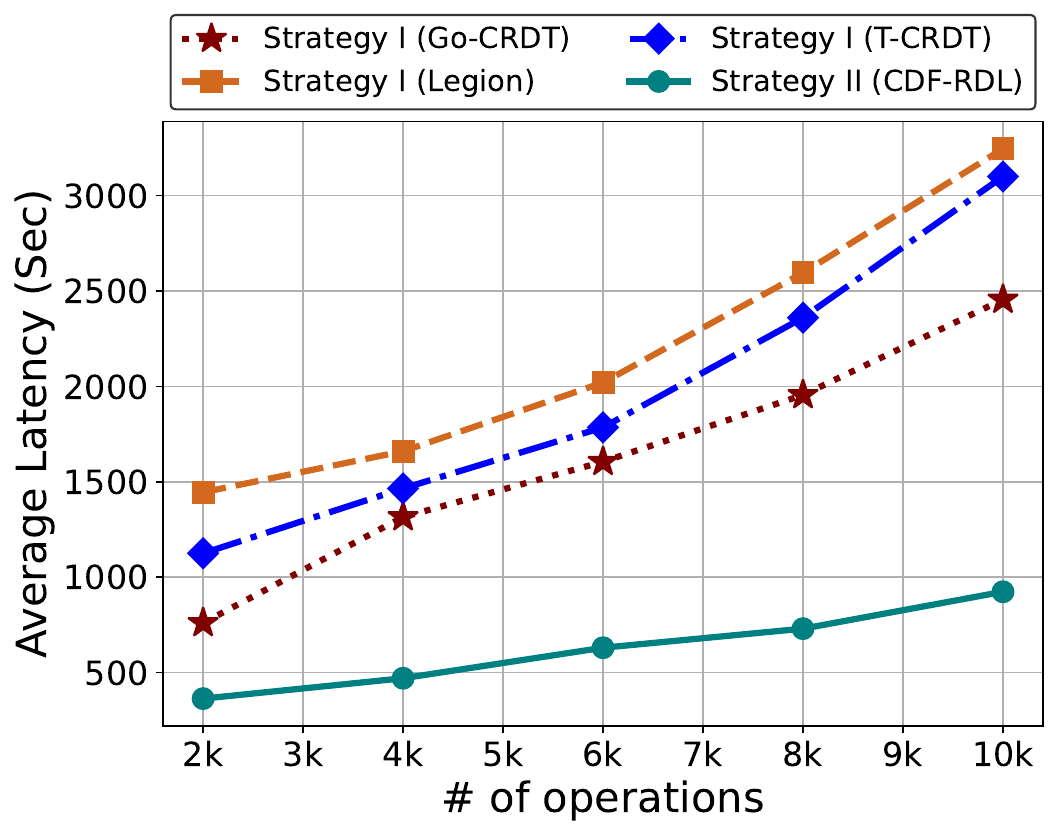}}\hfill
    \subfloat[Set\label{subfig:set_latency}]
    {\includegraphics[width=.33\linewidth]{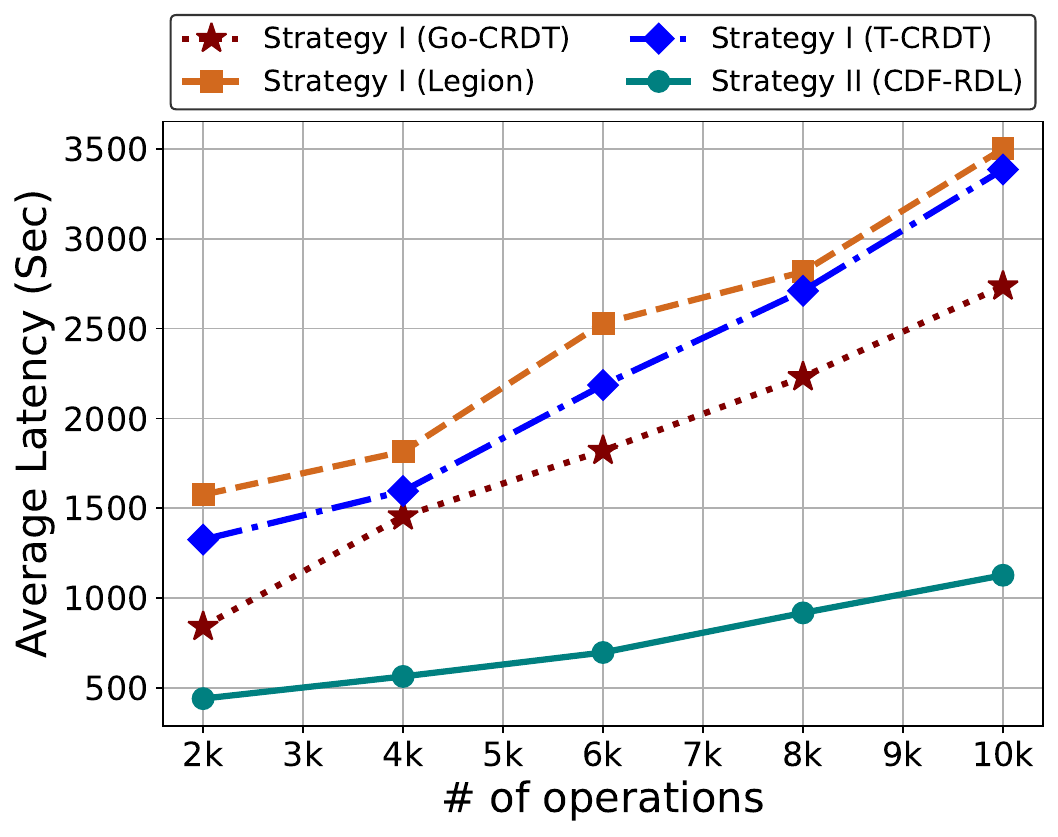}}\hfill
    \subfloat[Map\label{subfig:map_latency}]
    {\includegraphics[width=.33\linewidth]{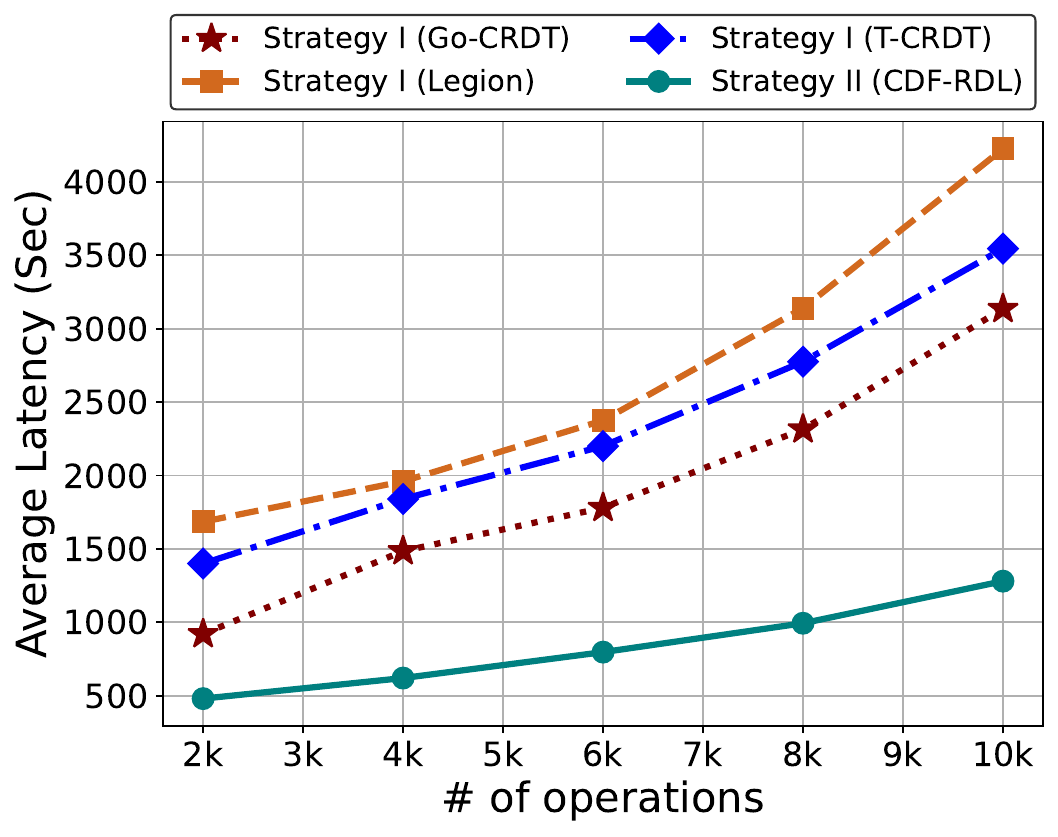}}\hfill
    \caption{Average Latency for Strategy I and II RDLs}
    \label{fig:latency}
\end{figure*}

\begin{figure*}[hbt!]
    \centering
    \subfloat[Counter\label{subfig:counter_memory}]
    {\includegraphics[width=.33\linewidth]{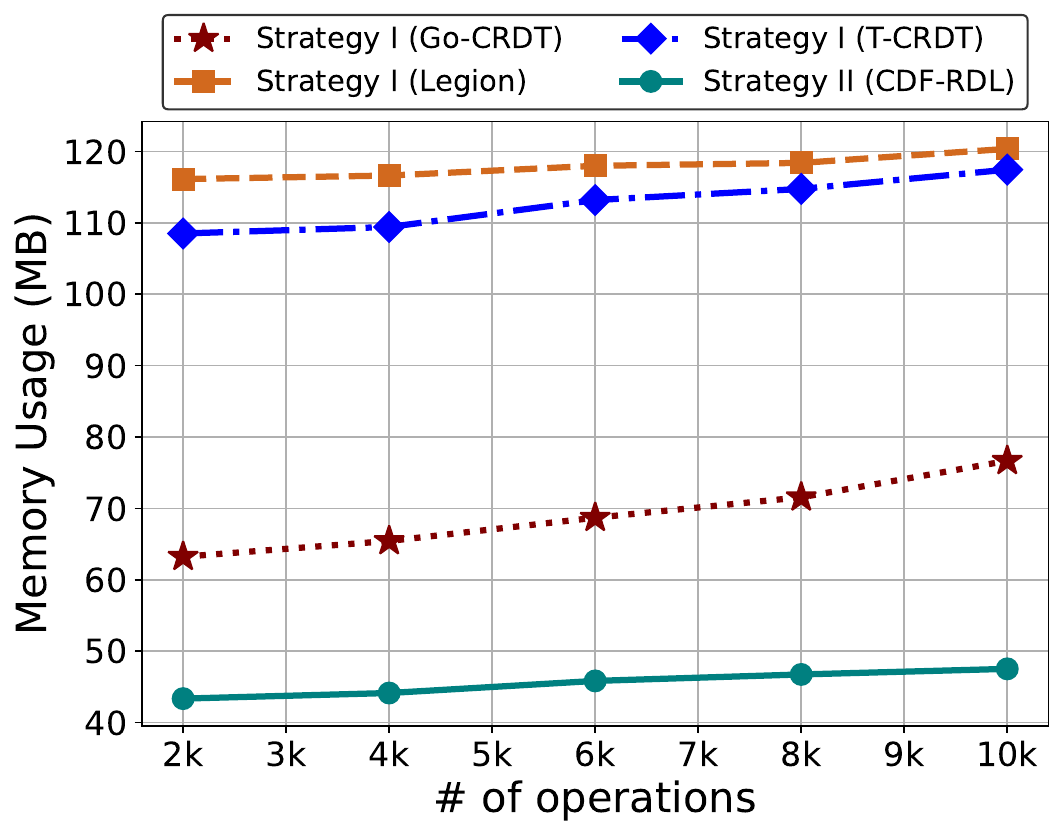}}\hfill
    \subfloat[Set\label{subfig:set_memory}]
    {\includegraphics[width=.33\linewidth]{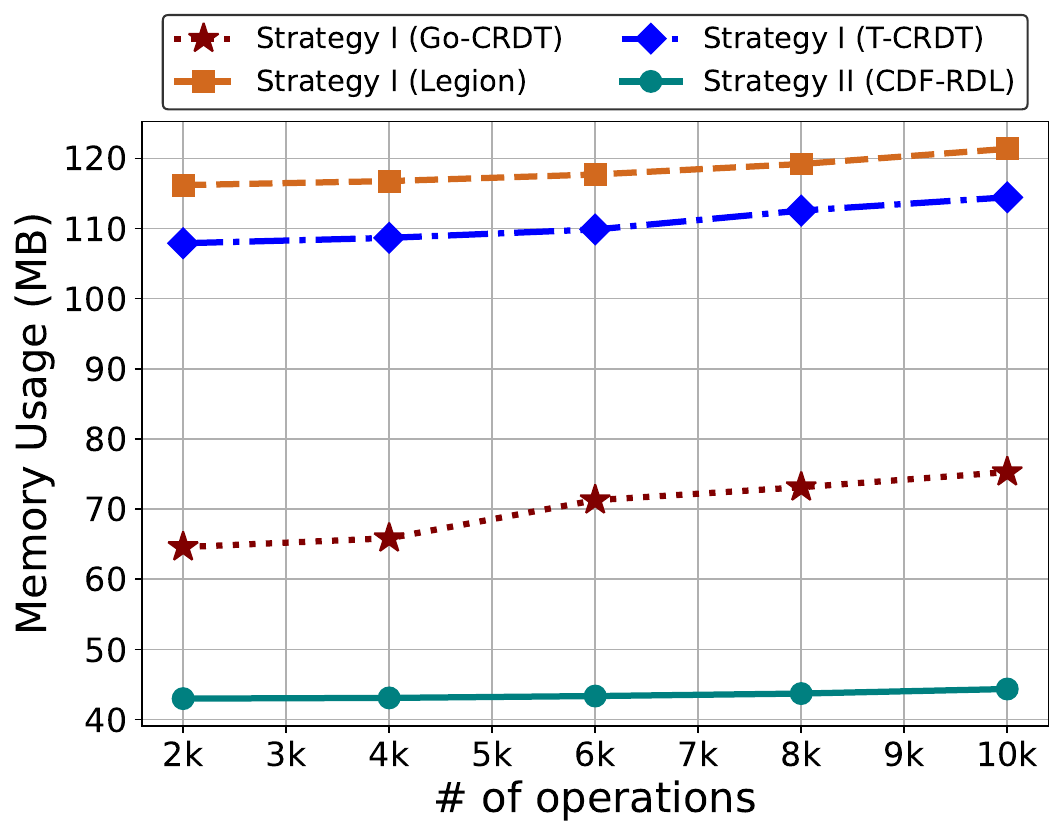}}\hfill
    \subfloat[Map\label{subfig:map_memory}]
    {\includegraphics[width=.33\linewidth]{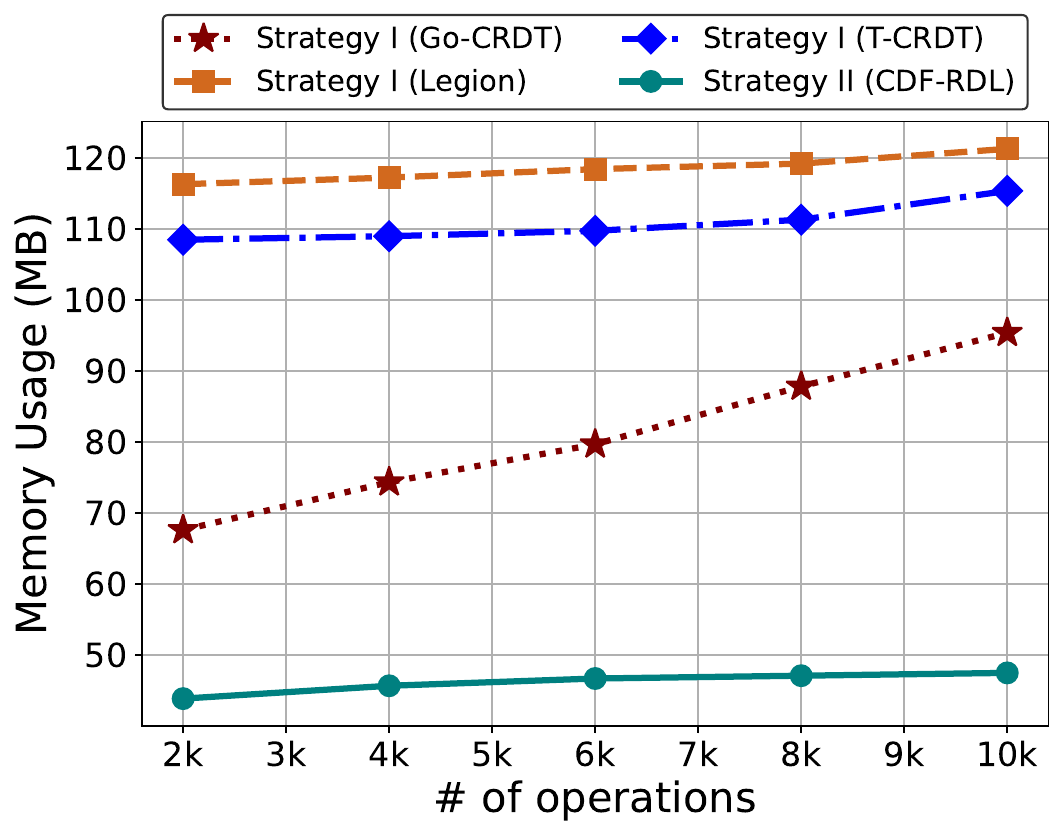}}\hfill
    \caption{Average Peak Memory Usage for Strategy I and II RDLs}
    \label{fig:memory}
\end{figure*}

\begin{figure*}[hbt!]
    \centering
    \subfloat[Counter\label{subfig:counter_throughput}]
    {\includegraphics[width=.33\linewidth]{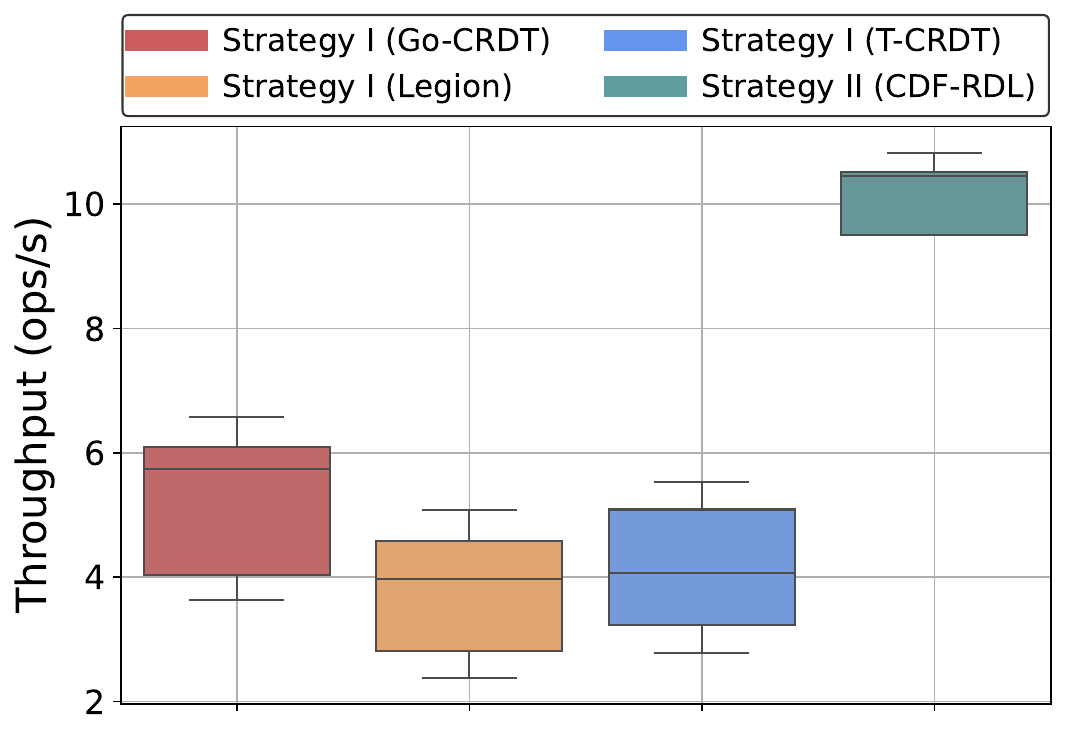}}\hfill
    \subfloat[Set\label{subfig:set_throughput}]
    {\includegraphics[width=.33\linewidth]{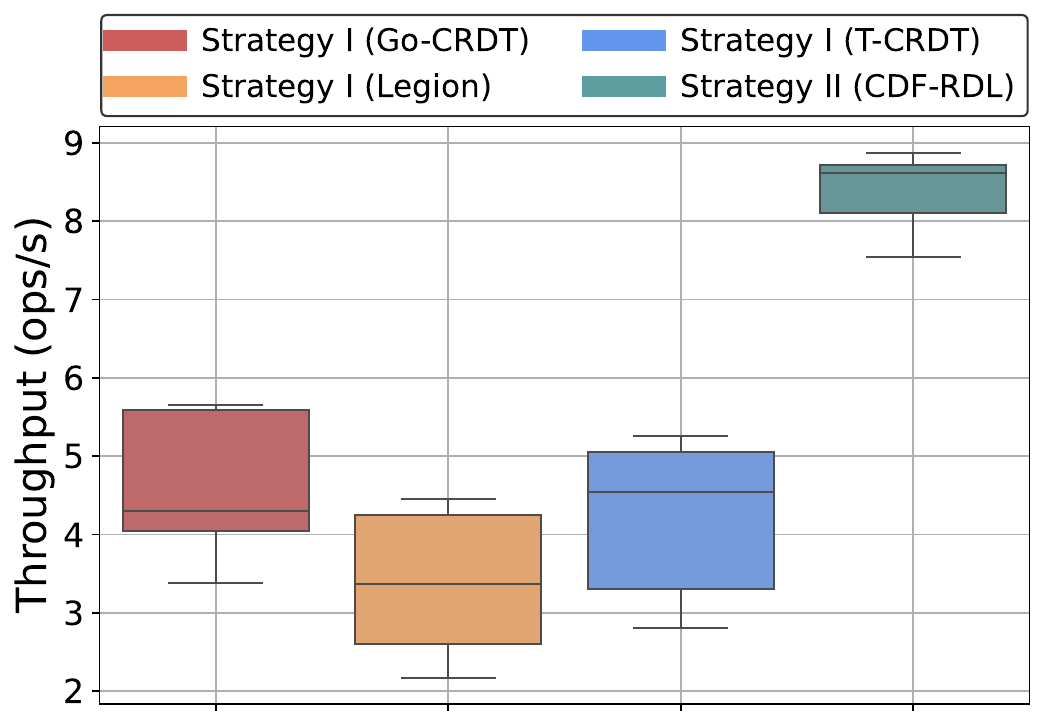}}\hfill
    \subfloat[Map\label{subfig:map_throughput}]
    {\includegraphics[width=.33\linewidth]{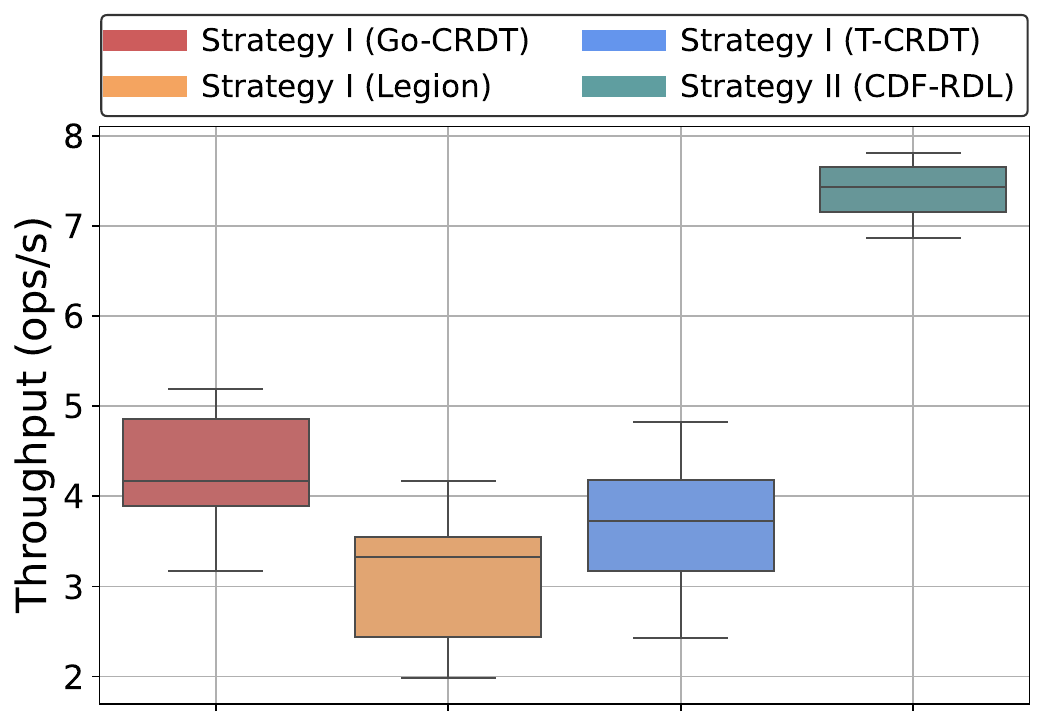}}\hfill
    \caption{Throughput for Strategy I and II RDLs}
    \label{fig:throughput}
\end{figure*}

To evaluate the performance characteristics of our evaluation subjects, we executed several benchmarks whose workloads differed by the number of RDL operations. The benchmarked operations concerned the fundamental RDL functionalities of accessing, updating, merging, and propagating the replicated states. We excluded the operation that collects ambient data in the IoT device, as this functionality is not part of RDL. Figures ~\ref{subfig:counter_latency}, ~\ref{subfig:set_latency}, and ~\ref{subfig:map_latency} depict the average latency of Counter, Set, and Map, respectively. We report the latency values measured for Strategy I's Go-CRDT, Legion, and T-CRDT and Strategy II (CDF-RDL). Across all measurements, the latency increased proportionally to the workload. However, CDF-RDL exhibited the lowest latency characteristics as compared to the other three evaluation subjects. This efficiency is due to CDF-RDL's capability to interact with the application logic in the same languages, whereas the other RDLs rely on FFI for such interactions. Among the other three evaluation subjects, Legion showed the worst performance due to the heavy overhead of JavaScript's runtime and interaction mechanisms, which involve WebAssembly as the FFI enabler. The need for compiling code at runtime incurs a high level of performance overhead, not present in the interactions between Go and Java, which require no runtime code generation~\cite{sestoft2002runtime}. Among the T-CRDT and Go-CRDT, the Go-CRDT takes less time due to Go compiling to native code. On average, CDF-RDL showed the shortest latency, compared to Go-CRDT, Legion, and T-CRDT by $\approx2.54\times$, $\approx3.46\times$, and $\approx3.11\times$, respectively. 

Figures ~\ref{subfig:counter_memory}, ~\ref{subfig:set_memory}, and ~\ref{subfig:map_memory} depict the average of the peak memory usage by Counter, Set, and Map, respectively. For each of the three data types, memory consumption increases slightly with the number of operations in the workload. However, for Set and Map, memory consumption increases more than that of Counter, as the size of these two data types impacts their memory consumption to a higher degree. CDF-RDL showed the smallest memory consumption due to the efficiency of the application code invoking RDL functions in the same language. On average, CDF-RDL outperformed Go-CRDT, Legion, and T-CRDT by $\approx1.63\times$, $\approx2.64\times$, and $\approx2.47\times$, respectively. 

Figures ~\ref{subfig:counter_throughput}, ~\ref{subfig:set_throughput}, and ~\ref{subfig:map_throughput} demonstrate the throughput distribution for Counter, Set, and Map, respectively. CDF-RDL's median throughput is the highest, indicating its superior performance as compared to the other evaluation subjects under the same workload. CDF-RDL's interquartile range (IQR) is always the narrowest due to the diminished variability across workloads. In contrast, the high IQR of the other subjects indicates a high variability of their throughput due to their reliance on dissimilar FFI techniques across languages. 

\paragraph*{FFI Implementation Insights}
To better understand why using FFI increases the programming burden and code complexity, we next shed light on some of the specific techniques we used to enable different languages to interact with each other over FFIs. As it turns out, multilingual functionalities are far from trivial to implement and maintain in the modern development ecosystem. For Go$\rightarrow$Java, we used \texttt{Java Native Interface (JNI)}, which allows Go code to invoke Java methods through native C functions. However, to be able to invoke the JNI C functions, the Go code had to use the special-purpose \texttt{cgo} library~\cite{metal3d2024} with its domain-specific usage conventions. For Java$\rightarrow$Go, the Go functions had to be wrapped in the C-calling-convention functions so they can be exported and be accessible via the \texttt{Java Native Access (JNA)} API. For Go$\rightarrow$JavaScript, we used the \texttt{go/goja}~\cite{goja_github} package, which provides an ECMAScript 5.1 JavaScript runtime within Go, through which Go code can invoke JavaScript functions. For JavaScript$\rightarrow$Go, we used \texttt{syscall/js}~\cite{syscall_js_godoc} library that makes it possible to compile Go functions to the WebAssembly format, which can then be invoked from JavaScript. Finally, for Java$\leftrightarrow$JavaScript, we used \texttt{Nashorn Engine}~\cite{sharan2014java}, which embeds a JavaScript engine within Java, through which Java code can invoke JavaScript functions and vice versa.
 
Combining different languages via FFIs, we encountered great variability across techniques and frameworks. We were also struck by how fragile some of these technologies are. Many of them require that a particular language version or a set of features be used, with several crippling limitations. For example, the Nashorn engine used for interactions between Java and JavaScript provides no support for modern ES6 features like Set, Map, Arrays, etc. One might argue with the FFI choices we have made, but our conclusions are likely to stand with possible alternatives. All FFI technologies we have used have turned error-prone, fragile, and required a steep learning curve.

\paragraph*{Threats to Validity} The results of our empirical study are subject to several threats to validity and limitations. First, our experimental results are derived from our example application, which may have influenced some of our design decisions. Nevertheless, the three evaluated RDLs were third-party, and we used their APIs to support our example application. Our example application exhibits all features of a multilingual replicated data system, representative of similar systems. Unfortunately, commercial systems of this type are unlikely to be open-sourced, making them inaccessible for experimentation. The ones that are open-sourced lack sufficient complexity in terms of mixing multiple languages to make viable experimental subjects. Another threat to validity is our choice of specific FFI libraries. The FFI technology space is quite diverse, with several competing libraries and frameworks. Nonetheless, we opted for using the most common exemplars for each language pair. Finally, our Strategy II relies on protobuf, thus inheriting its performance characteristics for each supported language. Hence, our measurements can be skewed by how well the protobuf libraries are optimized for each evaluated language.

\subsection{Discussion}
Various factors influence the choice of an implementation language. Sometimes, domain constraints require using a particular language or language type. For example, the resource constraints of edge computing may necessitate selecting a language that minimizes resource utilization~\cite{maheswaran2019language}. Some languages require extensive runtime environments, such as virtual machines or interpreters, which may be impractical to install on resource-limited devices like those in IoT setups~\cite{li2018everylite}. Hence, modern distributed applications mix languages, and as such, need to rely on various FFI technologies and tools for cross-language interoperability~\cite{yang2024multi, grichi2020towards}. Additionally, various business scenarios require maintaining data replicas~\cite{goel2007data, cecchet2008middleware}. These applications thus would benefit from an RDL that supports multilingual integration. 

Our empirical study has demonstrated the advantages of using a multilingual RDL that relies on CDF for cross-replica coordination over alternatives that use a monolingual RDL with FFI bindings to each language. Having compared their respective software and performance quality characteristics, we derived actionable insights for the design and implementation of RDLs for multilingual replicated data systems. To further validate our findings and capitalize on the derived insights, we put them into practice by creating \tool.

In addition, it would be impossible to create an RDL that provides \emph{all} functionalities required in different application scenarios. For instance, consider error handling in our motivating example~\ref{sub:motivating-example}. \texttt{Replica Go} could collect faulty sensor data (e.g., incorrect temperature values due to hardware malfunction); \texttt{Replica Java} could incorrectly persist the data without proper validation, while the Node.js frontend of \texttt{Replica JS} could display inaccurate visualizations. If an RDL lacks error-handling support, application programmers would need to implement this additional functionality in the application space. Providing effective error handling for a distributed system is challenging~\cite{armstrong2003making}. Doing so in a multilingual environment would be even harder due to the heterogeneity of languages, frameworks, and data exchange formats. If an RDL does provide error handling, utilizing it across multilingual replicas would face challenges similar to those of using the library itself. Hence, the issue of extensibility goes hand-in-hand with the integration strategies of RDLs in multilingual environments. We further validate our study's findings by making \tool plug-in extensible. 

\section{System Overview} \label{sec:system}
\begin{figure*}[hbt]
    \center
    \includegraphics[width=0.8\textwidth]{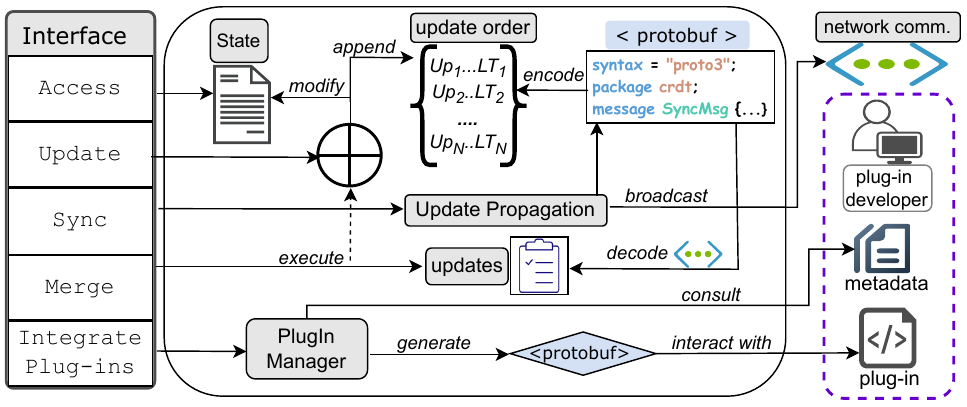}
    \vspace{-0.05in}
    \caption{\tool's System Overview and Workflow}
    \vspace{-1.0em}
    \label{fig:system}
    \centering
\end{figure*}

In this section, we describe \tool's system assumptions, design, and the workflow between its components. 

\subsection{System Assumptions} 
\tool's design assumes a replicated data system comprising a set of uniquely identified replicas. These replicas communicate via asynchronous messages. The system may experience a replica to crash, with the ability to completely recover, restoring the replica’s memory in its original form from durable storage space. The connecting network can recover from temporary partitions and also ensures reliable message delivery, not losing, distorting, or multiplying the messages. We assume that all participating replicas cannot fail at the same time.

\subsection{System Components and Workflow}
Figure \ref{fig:system} depicts \tool's system components and their interactions. \tool's Interface component provides five API functions for application developers: (i) \texttt{Access}, (ii) \texttt{Update}, (iii) \texttt{Sync}, (iv) \texttt{Merge}, and (v) \texttt{Integrate Plug-ins}. The functions from (i) to (iv) are commonly found across various RDLs; in contrast, function (v) is unique to \tool, responsible for its plug-in extensibility. 

Following classic RDL design, the core API functions deliver the following functionalities. \texttt{Access} retrieves the local replica's state. \texttt{Update} modifies the local replica's state in a fashion required by the data structure (e.g., increment/decrement for Counter, or add/remove for Set, etc.). When \texttt{Update} modifies the local state, the replica's \emph{update order}---a replica's causal record of update operations---is modified accordingly. The \emph{update order} is represented with Lamport timestamps (LT)~\cite{lamport2007lamport}. \texttt{Update} can be invoked by both the application code and the RDL implementation itself to process the merge requests from other replicas. 


\texttt{Sync} propagates a replica's local state to the remaining replicas. As \tool's \emph{raison d\textquotesingle\^etre} is to support cross-language interaction across multilingual replicas, its Update Propagation component provides enhanced services. It can take a propagation request in the format of any supported source language and deliver it in the format of the target language. To that end, \tool represents the update order in a language-independent fashion. The reference implementation uses the CDF provided by Google Protocol Buffer (protobuf) as a mature framework. With the help of protobuf, \tool encodes the update order and then broadcasts the encoded message over the network, marked as \emph{network comm.} in the diagram. \texttt{Merge} gets triggered when a replica receives synchronization requests from other replicas. To execute the merge, \tool first decodes the arrived message via protobuf, thus extracting a list of updates. Then \tool executes the updates on the list one by one by invoking the \texttt{Update} function. 

Due to the great diversity of application domains and types of applications that use replicated data, it would be impossible to determine a complete set of features that should be provided by an RDL, particularly if it is multilingual. To that end, \texttt{Integrate Plug-ins} provides a systematic avenue for extending \tool with additional features by following the plug-in architectural pattern. \tool's programming model for developing third-party plug-ins includes two affordances: (1) receiving the information (e.g., parameters, state changes, etc.) of the invoked core functions; (2) requesting the invocation of any core function. Notice that these affordances embody bi-directional interaction between \tool's core API and plug-ins.

To configure which plug-in to execute, developers can parameterize the \texttt{Integrate Plug-ins} function with metadata that contains the information about how the plug-in interacts with RDL core functions. Plug-ins are differentiated by unique system-wide IDs, also passed as parameters. \texttt{Integrate Plug-ins} uses the PlugIn Manager component that, having consulted metadata, generates the protobuf code required for the interaction between the plug-in and RDL core functions. By relying on CDF, \tool ensures cross-language interaction not only for core RDL functions but also for its plug-in extensibility. Supporting plug-ins written in a single language of the developer's choosing, \tool facilitates plug-in extensibility in a multilingual environment. \tool integrates these single-language plug-ins with core RDL functions, irrespective of which languages these functions might be implemented in. 
 
\section{Implementation} \label{sec:implementation}
Next, we describe \tool's reference implementation. Our description includes \tool's data structures and their implementation in different languages. We conclude the section by describing \tool's plug-in extensibility. 

\subsection{\tool's Data Structures}
Recall that \tool currently supports three RDL data structures: Counter, Set, and Map. A counter is a replicated integer and can be used in distributed systems to maintain a consistent count across replicas. For example, a replicated counter can be used for counting the number of likes across replicated social media sites, so each replica increments or decrements its local counter state, with the counters on each site eventually converging~\cite{almeida2019scalable}. A Set data type maintains a collection of unique elements across replicas. For example, a set can represent an online gift registry, where the same item can be added from multiple replicas, but only one copy ends up in the set~\cite{laddad2022keep}. A Map data type maintains a collection of key-value pairs across replicas. This data type is more expressive, as each of its keys can be associated with another data type (like Counter or Set) as their values. For example, a replicated map can maintain a shopping cart in an e-commerce application~\cite{de2012key}. Each cart product ID (key) can be associated with a Counter RDL (value) that tracks the quantity. 

\subsection{Multilingual Implementation \& Interoperability}
We aim to demonstrate \tool's versatility within a manageable implementation effort. Hence, our approach strategically selects one language from the three main execution models: compiled, interpreted, and managed. Specifically, \tool supports Go, JavaScript, and Java, respectively. By demonstrating the ability to support these dissimilar languages, we argue that by extending engineering effort, one can provide support for other languages that follow these execution models. In terms of the required engineering effort, it took us $\approx680$ lines of Go code, $\approx510$ lines of JavaScript code, and $\approx565$ lines of Java code to implement the aforementioned three RDL data structures. These ULOC measurements indicate the implementation effort required to provide the common RDL core functionalities of \texttt{Access}, \texttt{Update}, \texttt{Sync}, and \texttt{Merge}.

These are popular languages, so it is not surprising that they have been used to implement several existing RDLs. However, none of these libraries can be applied in a multilingual environment without providing customized bindings. Instead, these libraries have become a source from which \tool draws implementation insights on how to provide the required building blocks for constructing replicated data systems. What distinguishes our approach is its seamless cross-language replica coordination via CDF. We use the CDF of protobuf to define the \emph{update order}, propagated to the remaining replicas to ensure convergence. The protobuf libraries for each language\footnote{google.golang.org/protobuf~\cite{golang_protobuf} for Go, npm protocol-buffers~\cite{npm_protobuf} for JavaScript, and protobuf-java-3.21.12.jar~\cite{java_protobuf} for Java.} transform and propagate the transformed data across different replicas.

\subsection{\tool's Plug-in Extensibility}
\tool's \texttt{Integrate Plug-ins} API function performs two tasks: (1) consulting the provided metadata to determine the information that needs to be shared between the provided plug-in and \tool's PlugIn Manager component and (2) generating the corresponding protobuf code that reflects the information. To accomplish task (1), parameterized with a unique plug-in ID and corresponding metadata, this function checks for the presence of a JSON metadata file. For example, to integrate the plug-in whose ID is $n$, developers can invoke \lstinline{IntegratePlugIns(n, metaData_n.json)}. Based on the metadata, \tool then generates and compiles the protobuf code required for the interaction, thus accomplishing task (2). If the compilation fails, then \tool reports an error, abandoning the attempt to integrate that plug-in. If successfully compiled, \tool then deploys the plug-in to communicate with the core RDL functionalities via a socket, an implementation choice we have made. The language-specific parts of \texttt{Integrate Plug-ins} were implemented in $\approx90$, $\approx70$, and $\approx85$ ULOC of Go, JavaScript, and Java, respectively. Procedure~\ref{proc:plugin} demonstrates the key steps that \tool follows to integrate a plug-in.

\begin{algorithm}[hbt!]
    {\fontsize{9pt}{9pt}\selectfont
    \caption{Integrate Plug-ins Function}
    \label{proc:plugin}
    \begin{algorithmic}[1]
    \Procedure{Integrate\_PlugIns($pluginIDs, metadata$)}{}
        \ForAll{$pId \in pluginIDs$}
            \State $protoSyntax \gets$ consult$(metadata)$
            \State $protobufCode \gets$ generate$(protoSyntax)$
            \If {compile$(protobufCode)$}
                \State $pluginCode \gets$ locate$(pId)$
                \State $address \gets$ GetpluginAddress$(metadata)$
                \State deploy$(pluginCode, address)$
                \Statex \MyComment{Deploy $pluginCode$ at localhost:$address$}
                \State interact$(PlugInManager, pluginCode, address)$
            \Else
                \State reportError$()$
                \State \textbf{continue} \MyComment{proceed to next plug-in}
            \EndIf
        \EndFor
    \EndProcedure
    \end{algorithmic}
    }
\end{algorithm}

\subsection{Software Quality of \tool's Plug-in Extensibility} 
To validate \tool's mechanism for extending the core functions with additional functionalities, we conducted an experiment that entailed providing three non-trivial features. The provided features are (i) logging, (ii) undo, and (iii) rollback. Logging, a common middleware feature, provides a detailed and chronological record of all updates executed across distributed replicas~\cite{guo2017distributedlog}. Frequently used as an audit trail, this record enables developers to trace the sources of errors, monitor system behavior, and ensure compliance with applied policies. As such, logging is essential for non-trivial debugging, reproducing bugs, and handling faults. Both undo and rollback are examples of error-handling strategies, essential for ensuring the reliability of distributed systems. Both strategies aim to handle errors by reverting the replicated state to a previous point. Despite their shared objective, these strategies operate in distinctly different ways. While rollback simply brings the replicated state to a given prior persisted state~\cite{vassor2018checkpoint}, undo counteracts the effects of erroneous updates, thus offering a more fine-grained form of state restoration~\cite{yu2019generic}.

To prepare the evaluation subjects, we integrated the aforementioned features in two modes: (1) using \tool's plug-in architecture and (2) in an ad-hoc fashion for the other three evaluation subjects. While \tool can integrate any extra feature implemented in a single language, the rest require that the feature be implemented in the same language as the RDL itself. Hence, we provided all three features in Go for Go-CRDT, JavaScript for Legion, and Java for T-CRDT. In terms of the implementation effort, whether provided as a plug-in or in an ad-hoc fashion, each feature took approximately the same amount of source code, with the ULOC numbers ranging from about $\approx200$ for logging to more than $\approx350$ for undo.

\begin{table}[hbt!]
\centering
\caption{Software Quality for Plug-in Extensibility}
\resizebox{0.75\columnwidth}{!}{
\begin{tabular}{l|l|c|c}
    \hline
     Features & Subjects & LCOM & CF  \\ \hline 
    \multirow{4}{*}{Logging} & \tool & \textbf{6} & \textbf{0.33} \\ 
     & Go-CRDT & 11 & 0.68  \\  
     & Legion & 9 & 0.61  \\  
     & T-CRDT & 12 & 0.76 \\ \hline 
    \multirow{4}{*}{Undo} & \tool & \textbf{8} & \textbf{0.57} \\ 
     & Go-CRDT & 13 & 0.79  \\ 
     & Legion & 12 & 0.84  \\ 
     & T-CRDT & 13 & 0.89 \\ \hline
    \multirow{4}{*}{Rollback} & \tool & \textbf{8} & \textbf{0.5}  \\
     & Go-CRDT & 12 & 0.57  \\ 
     & Legion & 14 & 0.76  \\ 
     & T-CRDT & 12 & 0.69  \\ \hline
\end{tabular}}
\label{table:plugins_eval}
\end{table}

We aim to assess the suitability of a plug-in architecture for introducing extra features, so we selected those software metrics that are commonly used for measuring modularity, a key indicator of strong separation of concerns~\cite{sarkar2008metrics}. Specifically, we collected Lack of Cohesion in Methods (LCOM) and Coupling Factor (CF). As higher modularity streamlines comprehensibility and maintainability, lower LCOM indicates better modularity~\cite{hitz1995measuring}. As fewer inter-dependencies across modules, lower CF indicates enhanced maintainability, higher scalability, and streamlined testing~\cite{abreu1996design}. Table ~\ref{table:plugins_eval} shows these metrics for each evaluated subject. Both LCOM and CF values are lower for \tool, highlighting its superior software quality, attributed to its plug-in architecture. This quality is achieved by encapsulating additional functionalities within modular plug-ins, each designed with a clear single-entry access point for external clients.
 
\section{Related Work} \label{sec:related}
Our empirical study and \tool's design draws inspiration from multiple related fields of inquiry, including empirical study of system architectures, cross-language interoperability, and system extensibility, which we discuss in turn.

\subsubsection*{Empirical Study of System Architectures}
Researchers study system architectures to form a fact-based opinion on the state of and possible improvements to existing system solutions. Such studies typically synthesize and evaluate existing design approaches, identifying their strengths, limitations, and applicability across representative use cases~\cite{maranzano2005architecture}. In particular, a rich prior body of research focuses on reviewing system architecture. A systematic review of the information system aspects of system-of-system (SoS) architecture has pointed out a slower maturity progress as compared to other software engineering fields~\cite{klein2013systematic}. Another effort identifies seven critical perspectives in SoS architecture selection, including the conceptual framework, evaluation criteria, interdependency, uncertainty, autonomy, dynamic evolution, and computational method~\cite{fang2021system}. To uncover optimal strategies for applying service-oriented architecture (SOA) modeling, another review presents a taxonomy of existing approaches and their critical comparison~\cite{mohsin2018review}. Sometimes, empirical studies are conducted to identify the most optimal design options as part of creating novel systems. For example, to fully understand the performance characteristics of different proxying strategies, EdgStr studies them to transform a two-tier client-cloud application into a three-tier client-edge-cloud application~\cite{an2024edgstr}. Several prior works focus on reviewing enterprise systems, pointing out their critical challenges~\cite{da2011enterprise, kaisler2005enterprise, salleh2012cloud}, and discussing their success factors and implementation strategies~\cite{ali2017erp}. Despite these extensive research efforts that study various system design strategies, none evaluate middleware designs for breaking language barriers in multilingual replicated data systems.

\subsubsection*{Cross-Language Interoperability} 
As distributed systems have become increasingly complex and need to utilize heterogeneous architectures, it would be unrealistic to expect a single language to be the best choice for an entire system. Hence, mixing different languages is a critical requirement for modern distributed systems. The design challenge is not only to enable different languages to interoperate seamlessly but also to avoid complex inter-language interfaces, rigid coding conventions, and poor performance~\cite{chisnall2013challenge}. TruffleVM, a virtual machine-based runtime, executes and combines multiple programming languages, with the languages interoperating via FFI~\cite{grimmer2018cross}. Another VM-based approach aims for a simple and low-overhead FFI for Python software~\cite{kell2011virtual}. FIG generates FFI wrappers from annotations for interoperating with C/C++ code~\cite{reppy2006application}. Jeannie enables C/Java  interactions via JNI by mixing the constituent code parts in the same file~\cite{hirzel2007jeannie}. Code-interoperability mode treats JVM and Java as a unified programming platform for prototyping heterogeneous code that can execute in parallel on various hardware and from different languages~\cite{stratikopoulos2023cross}. Automerge CRDT library is implemented in Rust with bindings to JavaScript via WebAssembly~\cite{Automerge}. Unlike these related approaches, which facilitate cross-language interoperability for specific languages, \tool presents an architecture that allows mixing languages across replicas without relying on rigid FFI conventions and associated performance overhead.

\subsubsection*{System Extensibility}
System extensibility refers to a design quality that allows systems to adapt and grow without requiring significant modifications to the core architecture. If properly designed, extensibility promotes modularity, so developers can build and deploy extensions independently, thereby reducing development time and maintenance costs. Hence, a rich body of prior research focuses on systematic feature extensibility. One recent approach discusses the challenges of extending knowledge-based system development platforms and describes the obstacles that stand in the way of integrating external functionality via a service-oriented approach~\cite{pavlov2021towards}. Another work focuses on optimizing the software architecture by efficiently allocating tasks to nodes, thus enabling real-time distributed systems to offer extensibility~\cite{zhu2010optimizing}. Keris, a backward-compatible extension of Java, allows creating and linking modules to provide a type-safe and non-invasive extension of Java applications~\cite{zenger2002evolving}. Several prior works apply plug-in architectures for systematically adding features in domains that include collaborative work~\cite{grundy2002engineering}, debugging and testing~\cite{campos2012gzoltar, greiler2010understanding}, and adaptive web service composition~\cite{charfi2009plug}. Although we derive valuable design insights from these prior works, our approach's novelty lies in applying plug-in extensibility to the design of multilingual RDLs.

\section{Conclusion} \label{sec:conclusion}
We have studied two strategies for integrating RDL into multilingual replicated data systems. Our study results have identified the inefficiencies of modern FFI interfaces, leading us to a design that provides an RDL in multiple languages while coordinating their interaction via a common data format. To fully capitalize on our findings, we have created this design's reference implementation, \tool, which supports replicas implemented in compiled, interpreted, and managed languages, allowing them to interoperate seamlessly. An additional benefit of \tool's system design is systematic extensibility. Specifically, developers can add new features as plug-ins in a single language, albeit integrated with all replica languages. To the best of our knowledge, this work is the first to empirically study the software engineering strategies for mixing languages in replicated data systems and provide a multilingual RDL design that supports plug-in extensibility.

As possible future work directions, we plan to further explore the versatility of \tool's design by supporting additional languages. \tool's extensibility can be further explored by providing new features in different languages. The impact of common data format choice can also be further investigated, particularly in relation to cross-language replica coordination. 

Because modern developers are often confined to using particular languages for different tasks, multilingual systems have become a mainstay of the computing ecosystem. As distributed systems increasingly need to replicate data, this paper contributes new insights and novel designs for replicated data libraries.

\bibliographystyle{IEEEtran}
\bibliography{references}

\end{document}